\documentclass[pra,aps,twocolumn,showpacs]{revtex4}

\usepackage{amsmath}
\usepackage{bm}
\usepackage{graphicx}

\begin{document}

\title{Measurement-induced spin squeezing in a cavity}

\author{Hiroki Saito}
\author{Masahito Ueda}
\affiliation{Department of Physics, Tokyo Institute of Technology,
Tokyo 152-8551, Japan \\
and CREST, Japan Science and Technology Corporation (JST), Saitama
332-0012, Japan}

\date{\today}

\begin{abstract}
Spin squeezing of an ensemble of $N$ atoms in a high-Q cavity is shown to
be enhanced by continuous measurement of photons in the cavity.
A degree of spin squeezing proportional to $N^{-1/3}$ is attained as a
quasi-steady state over a broad set of initial states of the atoms and
photons.
\end{abstract}


\maketitle

\narrowtext

\section{Introduction}

Squeezed spin states (SSS)~\cite{Kitagawa}, in which quantum uncertainty
of the collective spin in a direction orthogonal to the mean spin vector
is suppressed below the standard quantum limit, have recently been
realized experimentally~\cite{Hald,Kuzmich00,Juls}.
In the SSS of atomic systems, entanglement between atoms is crucial for
suppressing fluctuations in collective atomic spin, and therefore such
systems are of potential interest in quantum information processing as
well as in atomic interferometry.

A number of methods to generate SSS in atomic systems have been proposed,
and these can be classified into several categories.
A few examples that are relevant to the subject of the present paper are:
(i) Squeezing is transferred between photons and atoms via their
interactions~\cite{SaitoL,Kuzmich97,SaitoA,Poulsen}.
Using this property, SSS can be realized in an atomic system by copying
the squeezing from the system of photons to the atoms.
This scheme has been demonstrated experimentally~\cite{Hald}.
(ii) A probe laser beam propagating through atoms establishes entanglement
between the atoms, with measurement of the probe laser inducing
state reduction in the atomic states, resulting in
SSS~\cite{Kuzmich98,Takahashi,Duan}.
This kind of measurement-induced-squeezing scheme has also been employed
in experiments~\cite{Kuzmich00,Juls}.
(iii) The interaction between atoms and photons confined in a high-Q
cavity entangles them, and states of both atoms and photons develop into
squeezed states~\cite{Wineland,Waka,SaitoL,SaitoA}.

In the method (iii) proposed by Wineland {\it et al.}~\cite{Wineland} and
by us~\cite{Waka,SaitoL,SaitoA}, photons prepared in a coherent state and
atoms prepared in a coherent spin state undergo interactions in a high-Q
cavity, and after a certain interaction time SSS is realized temporarily,
but not in a steady state.
Precise control over the initial state and interaction time is therefore
required.
In the present paper, a novel scheme to generate SSS is proposed that
combines methods (iii) and (ii) to circumvent the stringent requirements
of the precise control needed for method (iii).
SSS is thus obtained as a ``quasi-steady state'', regardless of
the initial state, by performing continuous measurement of photons in the
cavity.
This drastically reduces experimental difficulties in producing SSS.
Moreover, a degree of spin squeezing proportional to $N^{-1/3}$, where
$N$ is the number of atoms, can be achieved using this method, in contrast
to the $N^{-1/4}$ attainable by method (iii)~\cite{SaitoA}.

This paper is organized as follows.
Section~\ref{s:continuous} describes the system we consider and develops a
theory of continuous measurement of photon number in the cavity that is
being pumped and contains atoms to be squeezed.
Section~\ref{s:SSS} shows that the atomic state in the cavity develops
into an SSS by continuous measurement of photons in the cavity.
Section~\ref{s:conclusion} discusses possible experimental schemes and
concludes this paper.

\section{Continuous Measurement of Photon Number in a Pumped Cavity with
Atoms}
\label{s:continuous}

\subsection{Description of the system}

\begin{figure}[tb]
\includegraphics[width=8.4cm]{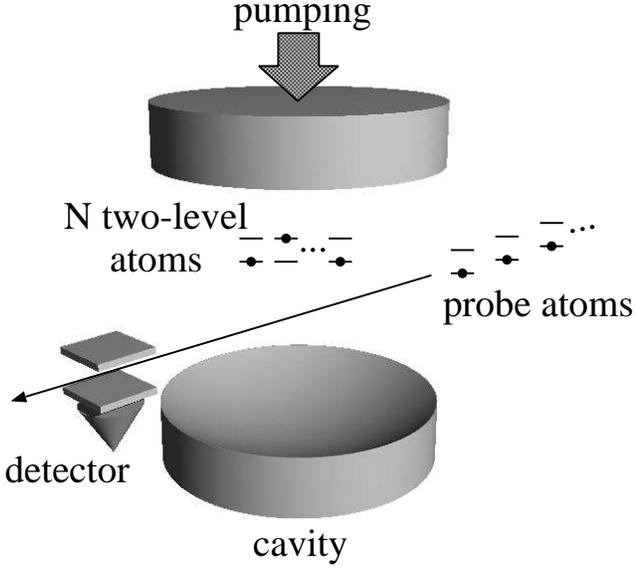}
\caption{
A schematic of the proposed system.
The $N$ two-level atoms interact with a single-mode photon field in a
pumped cavity.
The continuous measurement of photon number is simulated by a sequence of
ground-state probe atoms injected through the cavity, and the state of
each atom coming out of the cavity is measured by a detector.
If a probe atom is found to be in an excited (ground) state, we assume
that one photon (no photon) is detected.
}
\label{f:system}
\end{figure}

An operational theory of continuous photon counting has been proposed by
Srinivas and Davies~\cite{Srinivas}, and microscopic modeling of this
theory has been reported in Ref.~\cite{Imoto}.
The nonunitary state evolution due to measurement back action associated
with the Srinivas-Davies model was investigated in Ref.~\cite{Ueda}.
In these studies, continuous measurement of single-mode photons in a
high-Q cavity was considered.
We extend this model to the case of a pumped cavity containing $N$
two-level atoms as illustrated in Fig.~\ref{f:system}.

We first consider the system without photon counting, namely, without the
probe atoms and detector shown in Fig.~\ref{f:system}.
The high-Q cavity is assumed to sustain single-mode photons of frequency
$\omega_{\rm c}$ and be pumped by a coherent light source with an
effective amplitude $A$ and frequency $\omega_{\rm p}$ inside the cavity.
The cavity contains two-level atoms with transition frequency $\omega_{\rm
a}$, which couple to photons via the Jaynes-Cummings
interaction~\cite{JCM} with coupling constant $g$.
We assume that $g$ is constant and the same for all the atoms, that is, we
consider the Lamb-Dicke limit, where all atoms experience the same photon
field.
It is then convenient to introduce a set of collective spin operators
defined by
\begin{eqnarray}
\hat S_z & \equiv & \frac{1}{2} \sum_{j = 1}^N (|{\rm e}_j \rangle \langle
{\rm e}_j| - |{\rm g}_j \rangle \langle {\rm g}_j|), \\
\hat S_+ & \equiv & \frac{1}{2} \sum_{j = 1}^N |{\rm e}_j \rangle \langle
{\rm g}_j| \equiv \hat S_-^\dagger,
\end{eqnarray}
where $N$ is the number of atoms, and $|{\rm g}_j \rangle$ and $|{\rm e}_j
\rangle$ are the ground and excited states of the $j$th atom.
We shall also use other collective spin operators defined by $\hat S_x
\equiv (\hat S_+ + \hat S_-) / 2$ and $\hat S_y \equiv (\hat S_+ - \hat
S_-) / (2i)$.

The Hamiltonian for the system is given by
\begin{equation} \label{H}
\hat H_0 / \hbar = \omega_{\rm c} \hat n + \omega_{\rm a} \hat S_z + A
(e^{i \omega_{\rm p} t} \hat a + e^{-i \omega_{\rm p} t} \hat a^\dagger) 
+ g (\hat a \hat S_+ + \hat a^\dagger \hat S_-),
\end{equation}
where $\hat a$ ($\hat a^\dagger$) is the annihilation (creation) operator
of the photon field, $\hat n \equiv \hat a^\dagger \hat a$ is the
photon-number operator, and $A$ and $g$ are taken to be real and positive
without loss of generality.
In Eq.~(\ref{H}), the first and the second terms on the right-hand side
(rhs) describe the free Hamiltonian for the photon field and the
collective atomic system, respectively; the third term describes the
effect of coherent pumping of the photon field, and the last term
describes the interaction between the photon field and the atomic system.
In the rotating frame of reference $e^{i \omega_{\rm p} (\hat n + \hat
S_z)} |\psi \rangle$, the Hamiltonian (\ref{H}) is time-independent:
\begin{equation} \label{Hrot}
\hat H_0^{\rm rot} / \hbar = \delta_{\rm cp} \hat n + \delta_{\rm ap} \hat
S_z + A (\hat a + \hat a^\dagger) + g (\hat a \hat S_+ + \hat a^\dagger
\hat S_-),
\end{equation}
where $\delta_{\rm cp} \equiv \omega_{\rm c} - \omega_{\rm p}$ and
$\delta_{\rm ap} \equiv \omega_{\rm a} - \omega_{\rm p}$.

We now consider continuous photon counting for the system described by the
Hamiltonian (\ref{Hrot}).
Experimentally, photon counting may be performed using a photodetector
placed in the immediate vicinity of the cavity, where photons leaking out
of the cavity are detected.
This kind of continuous photodetection can be simulated using a sequence
of probe atoms that are injected through the cavity~\cite{Imoto}, as
illustrated in Fig.~\ref{f:system}.
The probe atoms are prepared in the ground state and injected into the
cavity, with internal states detected upon leaving the cavity.
If a probe atom is found to be in an excited state, we can say that one
photon has been detected ({\it one-count process}).
If it is found to have remained in the ground state, we can say that no
photon has been detected ({\it no-count process})~\cite{Imoto,Ueda}.
In the remaining part of this section we discuss nonunitary state
evolution of the system in the one-count process, the no-count process,
and more general processes.

\subsection{State evolution in the one-count process}

If the passage time of each probe atom through the cavity is much shorter
than the time scale of the state evolution described by Eq.~(\ref{Hrot}),
the density operator of the entire system (i.e., cavity photons plus atoms
excluding the probe atoms) immediately after the one-count process at time
$t$ is given by (see Appendix~\ref{a:continuous} for the derivation)
\begin{equation} \label{onecount}
\hat\rho(t_+) = \frac{\hat a \hat\rho(t) \hat a^\dagger}{{\rm Tr}[\hat
a \hat\rho(t) \hat a^\dagger]},
\end{equation}
where $t_+$ denotes a time infinitesimally later than $t$ and ${\rm
Tr}$ the trace of the entire system.
The denominator on the rhs of Eq.~(\ref{onecount}) multiplied by $\gamma
\Delta t$ gives the probability that a photon is detected between $t$ and
$t + \Delta t$, where $\gamma$ is the photodetection rate defined by
Eq.~(\ref{gamma}).
The expression (\ref{onecount}) takes the same form as that of the
Srinivas-Davies model~\cite{Srinivas}.
Immediately after the one-count process (\ref{onecount}), the expectation
value of an arbitrary operator $\hat {\cal O}$ is given by
\begin{equation} \label{onecountO}
{\rm Tr}[\rho(t_+) \hat {\cal O}] \equiv \langle \hat {\cal O}
\rangle_{t_+} = \frac{\langle \hat a^\dagger \hat {\cal O} \hat a
\rangle_t}{\langle \hat n \rangle_t}.
\end{equation}

When $\hat {\cal O}$ is the photon-number operator $\hat n$, we obtain
\begin{equation} \label{onecountn}
\langle \hat n \rangle_{t_+} = \langle \hat n \rangle_t - 1 +
\frac{\langle (\Delta \hat n)^2 \rangle_t}{\langle \hat n \rangle_t},
\end{equation}
where $\Delta \hat n \equiv \hat n - \langle \hat n \rangle$.
The one-count process thus decreases the mean photon number by one and
increases it by the Fano factor $\langle (\Delta \hat n)^2 \rangle_t /
\langle \hat n \rangle_t$ of the photon field immediately before the
one-count process~\cite{Ueda}.

When $\hat {\cal O}$ is some spin operator $\hat {\cal S}$, which commutes
with $\hat a$, Eq.~(\ref{onecountO}) becomes
\begin{equation}
\langle \hat {\cal S} \rangle_{t_+} = \frac{\langle \hat {\cal S} \hat n
\rangle_t}{\langle \hat n \rangle_t} = \langle \hat {\cal S} \rangle_t +
\frac{\langle \Delta \hat {\cal S} \Delta \hat n \rangle_t}{\langle \hat n
\rangle_t}.
\end{equation}
This indicates that upon the detection of a photon, the expectation value
of the collective spin operator changes by an amount that depends on the
correlation between the photons and atoms.
As an example, let us consider an entangled state between an atom and a
photon, where the ground and excited states of the atom are denoted by
$|{\rm g} \rangle$ and $|{\rm e} \rangle$, and the vacuum and
single-photon state by $|0 \rangle$ and $|1 \rangle$.
Suppose that the state of the system before measurement is given by
$(|{\rm g} \rangle |1 \rangle + |{\rm e} \rangle |0 \rangle) / \sqrt{2}$.
Then $\langle \hat S_z \rangle$ decreases from $1/2$ to $0$ by the
one-count process.
This is because the measurement reveals that the cavity had contained a
photon, and hence that the state of the system after the one-count process
is given by $|{\rm g} \rangle |0 \rangle$.
Similarly, $\langle \hat S_z \rangle$ increases upon the one-count process
from $1/2$ to $1$ for an entangled initial state $(|{\rm g} \rangle |0
\rangle + |{\rm e} \rangle |1 \rangle) / \sqrt{2}$.
In contrast, when the initial state is separable, i.e., $\langle \hat
{\cal S} \hat n \rangle_t = \langle \hat {\cal S} \rangle_t \langle \hat n
\rangle_t$, the atomic quantities do not change by the one-count process.

\subsection{State evolution in the no-count process}

When no photon is detected between $t$ and $t + T$, the density operator
is shown to evolve according to the following equation (see
Appendix~\ref{a:continuous} for the derivation),
\begin{equation} \label{nocount2}
\hat\rho(t + T) = \frac{e^{-i \hat H_{\rm nc} T / \hbar} \hat\rho(t) e^{i
\hat H_{\rm nc}^\dagger T / \hbar}}{{\rm Tr}[e^{-i \hat H_{\rm nc} T /
\hbar} \hat\rho(t) e^{i \hat H_{\rm nc}^\dagger T / \hbar}]},
\end{equation}
where $\hat H_{\rm nc} \equiv \hat H_0^{\rm rot} - i \hbar \gamma \hat n /
2$ is the non-Hermitian operator that governs the state evolution during
the no-count process.
The denominator on the rhs of Eq.~(\ref{nocount2}) gives the probability
that no photon is detected between $t$ and $t + T$.
It follows from Eq.~(\ref{nocount2}) that the master equation for the
no-count process is given by
\begin{equation} \label{nocount1}
\frac{\partial \hat \rho(t)}{\partial t} = \frac{i}{\hbar} [ \hat \rho(t)
\hat H_{\rm nc}^\dagger - \hat H_{\rm nc} \hat \rho(t) ] + \gamma \langle
\hat n \rangle_t \hat \rho(t).
\end{equation}
In fact, it can be shown by direct substitution of Eq.~(\ref{nocount2})
into Eq.~(\ref{nocount1}) that Eq.~(\ref{nocount2}) is the solution of
Eq.~(\ref{nocount1}).

\subsection{State evolution in the referring and non-referring measurement
processes}

Time evolution under continuous photodetection is thus a stochastic
process, in which the state evolution associated with one-count and
no-count processes are described by Eqs.~(\ref{onecount}) and
(\ref{nocount2}), respectively.
The probability of the one-count process occurring between $t$ and $t +
dt$ is $\gamma \langle \hat n \rangle_t dt$ and that of the no-count
process is $1 - \gamma \langle \hat n \rangle_t dt$.
The number of photons detected and the times at which they are detected,
therefore, differ run by run even for the same initial state.
The measurement process in which we read the outcome of the measurement
at every instant of time is referred to as {\it referring measurement
process}~\cite{Ueda}.
When the photons are detected at $t_1, t_2, \cdots, t_n$ from $t = 0$ to
$t = T$, the time evolution of the density operator is given by
\begin{equation}
\hat \rho(T) = \frac{\hat P(T; t_1, \cdots, t_n) \hat \rho(0) \hat
P^\dagger(T; t_1, \cdots, t_n)}{{\rm Tr}[\hat P(T; t_1, \cdots, t_n) \hat
\rho(0) \hat P^\dagger(T; t_1, \cdots, t_n)]},
\end{equation}
where
\begin{eqnarray} \label{refP}
\hat P(T; t_1, \cdots, t_n) & \equiv & e^{-i \hat H_{\rm nc} (T - t_n) /
\hbar} \hat a e^{-i \hat H_{\rm nc} (t_n - t_{n - 1}) / \hbar} \hat a
\nonumber \\
& & \cdots \hat a e^{-i \hat H_{\rm nc} (t_2 - t_1) / \hbar} \hat a e^{-i
\hat H_{\rm nc} t_1 / \hbar}.
\end{eqnarray}

If the detector is switched on but we do not read out the measurement
results, the state evolves into a statistical mixture of an ensemble of
all possible outcomes, which is referred to as {\it non-referring
measurement process}~\cite{Ueda}.
The master equation describing the state evolution in the non-referring
measurement is obtained as follows.
In the non-referring measurement process, $\hat \rho(t + dt)$ is a
statistical mixture of the density operator $\hat \rho_1(t + dt)$ of the
one-count process and the density operator $\hat \rho_0(t + dt)$ of the
no-count process:
\begin{equation} \label{nref1}
\hat \rho(t + dt) = \gamma \langle \hat n \rangle_t \hat \rho_1(t + dt) dt
+ \left( 1 - \gamma \langle \hat n \rangle_t dt \right) \hat \rho_0(t +
dt).
\end{equation}
Substituting $\hat \rho_1(t + dt)$ for $\hat \rho(t_+)$ in
Eq.~(\ref{onecount}) gives
\begin{equation} \label{nref2}
\gamma \langle \hat n \rangle_t \hat \rho_1(t + dt) dt = \gamma \hat a
\hat \rho(t) \hat a^\dagger dt.
\end{equation}
Setting $T = dt$ in Eq.~(\ref{nocount2}) gives
\begin{equation} \label{nref3}
\hat \rho_0(t + dt) = \hat \rho(t) + \frac{i}{\hbar} [ \hat \rho(t)
\hat H_{\rm nc}^\dagger - \hat H_{\rm nc} \hat \rho(t) ] dt + \gamma
\langle \hat n \rangle_t \hat \rho(t) dt + O(dt^2).
\end{equation}
Substituting Eqs.~(\ref{nref2}) and (\ref{nref3}) into Eq.~(\ref{nref1})
gives the master equation for the non-referring measurement:
\begin{equation} \label{master}
\frac{\partial \hat\rho(t)}{\partial t} = \frac{i}{\hbar} [\hat\rho(t),
\hat H_0^{\rm rot}] + \frac{\gamma}{2} \left[ 2 \hat a \hat\rho(t) \hat
a^\dagger - \hat n \hat\rho(t) - \hat\rho(t) \hat n \right].
\end{equation}

When the cavity contains no atoms or when $g = 0$, we can show that the
state of the photon field eventually evolves into a coherent state
$|\alpha \rangle$ with $\alpha = -A / (\delta_{\rm cp} - i \gamma / 2)$ in
both referring and non-referring measurement processes, unless the initial
state is orthogonal with $|\alpha^* \rangle$ (see Appendix~\ref{a:pump}).
This fact is interesting in that the photon number never increases for the
steady state $|\alpha \rangle$ despite the fact that the cavity is being
pumped and that no photons are detected.
This can be understood from the equation describing the time evolution of
the mean photon number in the no-count process
\begin{equation} \label{nev}
\frac{d \langle \hat n \rangle_t}{dt} = i A \langle \hat a - \hat
a^\dagger \rangle_t - \gamma \langle (\Delta \hat n)^2 \rangle_t,
\end{equation}
which can be obtained from Eq.~(\ref{nocount1}).
For the steady state $|\alpha \rangle$ with $\alpha = -A / (\delta_{\rm
cp} - i \gamma / 2)$, the pumping term [the first term of the rhs of
Eq.~(\ref{nev})] balances with the last term in Eq.~(\ref{nev}) which
describes the decrease in the number of photons due to the state reduction
by the no-count process.

\section{Generation of Squeezed Spin State by Continuous Photodetection}
\label{s:SSS}

\subsection{Squeezed spin state}

Because the spin operators $\hat S_j$ ($j = x, y, z$) obey the commutation
relation $[\hat S_x, \hat S_y] = i \hat S_x$ and the corresponding cyclic
permutations, the spin vector fluctuates around the mean value $\langle
\hat{\bf S} \rangle = (\langle \hat S_x \rangle, \langle \hat S_y \rangle,
\langle \hat S_z \rangle)$.
For example, let us consider the lowest eigenstate of $\hat S_z$, $|S, -S
\rangle$, where $S (S + 1)$ is the eigenvalue of $\hat{\bf S}^2 \equiv
\hat S_x^2 + \hat S_y^2 + \hat S_z^2$.
The mean spin vector of this state is given by $\langle \hat{\bf S}
\rangle = (0, 0, -S)$.
Quantum fluctuations around this are isotropic and given by $\langle
(\Delta \hat S_x)^2 \rangle = \langle (\Delta \hat S_y)^2 \rangle = S /
2$, satisfying the minimum uncertainty relation $\langle (\Delta \hat
S_x)^2 \rangle \langle (\Delta \hat S_y)^2 \rangle = \langle \hat S_z
\rangle^2 / 4$.
If the fluctuation of one component perpendicular to the mean spin vector,
say $\langle (\Delta \hat S_x)^2 \rangle$, is suppressed below $|\langle
\hat S_z \rangle| / 2$, the spin state is defined as
squeezed~\cite{Kitagawa}.

A general definition of spin squeezing is obtained by rotating the above
situation in spin space, when the mean spin vector $\langle \hat{\bf S}
\rangle$ points in an arbitrary direction.
The definition is thus given by~\cite{Kitagawa}
\begin{equation} \label{SSS}
\zeta \equiv \frac{2 \langle (\Delta \hat S_\perp)^2 \rangle}{|\langle
\hat{\bf S} \rangle|} < 1,
\end{equation}
where $\hat S_\perp$ is one spin component perpendicular to the mean spin
vector, $\Delta \hat S_\perp \equiv \hat S_\perp - \langle \hat S_\perp
\rangle$, and $|\langle \hat{\bf S} \rangle|^2 = \langle \hat S_x
\rangle^2 + \langle \hat S_y \rangle^2 + \langle \hat S_z \rangle^2$.
The squeezing parameter $\zeta$ is unity in the coherent spin state
\begin{equation} \label{css}
|\theta, \phi \rangle = \prod_j \left( \cos \frac{\theta}{2} |{\rm e}_j
\rangle + e^{i \phi} \sin \frac{\theta}{2} |{\rm g}_j \rangle \right).
\end{equation}
The definition (\ref{SSS}) of spin squeezing can also be applied to
situations in which the spin state is a mixed state.
The physical meaning of the SSS in atomic systems is discussed in
Ref.~\cite{SaitoA}.

For an $N$-atom system, there are $N / 2 + 1$ (when $N$ is even) or $(N +
1) / 2$ (when $N$ is odd) subspaces that can be indexed by the value of
the total spin $S = N / 2, N / 2 - 1, \cdots, 0 \: (1 / 2)$.
In the present paper, we restrict ourselves to the subspace of $S = N /
2$.
This corresponds to the case in which the initial state is the coherent
spin state, which is easily prepared in experiments.
Since the Hamiltonian (\ref{Hrot}) commutes with $\hat {\bf S}^2$, the
subspaces do not mix by time evolution.

\subsection{Analytic results}
\label{s:analytic}

For simplicity, we consider the case of $\delta_{\rm cp} = \delta_{\rm ap}
= 0$, i.e., the case in which both the coherent pumping field and the
atomic transition energy are resonant with the cavity mode.
The non-Hermitian operator $\hat H_{\rm nc}$ describing the no-count
process (\ref{nocount2}) is then given by
\begin{eqnarray} \label{atomnc}
\hat H_{\rm nc} / \hbar & = & A (\hat a + \hat a^\dagger) + g (\hat a \hat
S_+ + \hat a^\dagger \hat S_-) - \frac{i \gamma}{2} \hat n \nonumber \\
& \equiv & \hat a \hat \sigma^\dagger + \hat a^\dagger \hat \sigma -
\frac{i \gamma}{2} \hat n,
\end{eqnarray}
where we define $\hat \sigma \equiv g \hat S_- + A$.

As will be shown in Sec.~\ref{s:numerical}, the generation of SSS is
effective for $\gamma / g \gtrsim S$, and hence we derive an approximate
master equation of the no-count process for large $\gamma$.
In the following derivation, we assume $\gamma \gg A$ and $A \sim g S$.
Defining $\hat \rho'(t) \equiv e^{\gamma t \hat n / 2} \hat \rho(t)
e^{\gamma t \hat n / 2}$, we rewrite the master equation (\ref{nocount1})
for the no-count process as
\begin{eqnarray} \label{rhop}
\frac{\partial \hat \rho'(t)}{\partial t} & = & i \hat \rho'(t) \left(
 e^{\frac{\gamma t}{2}} \hat a \hat \sigma^\dagger + e^{-\frac{\gamma
 t}{2}} \hat a^\dagger \hat \sigma \right) \nonumber \\
& & - i \left( e^{-\frac{\gamma
 t}{2}} \hat a \hat \sigma^\dagger + e^{\frac{\gamma t}{2}} \hat a^\dagger
 \hat \sigma \right) \hat \rho'(t) + \gamma \langle \hat n \rangle_t \hat
 \rho'(t). \nonumber \\
\end{eqnarray}
Since $\gamma \gg A$, the photon field is almost always in the vacuum
state, which is justified below.
We may therefore approximate $\hat \rho'(t)$ as $\hat \rho'(t) \simeq
|0\rangle \langle 0| \otimes \hat \rho_s(t)$, where $\hat \rho_s(t)$
denotes the reduced density operator for the spin.
Integrating Eq.~(\ref{rhop}) iteratively gives
\begin{widetext}
\begin{eqnarray} \label{perturb}
\hat \rho'(t + \Delta t) & = & |0\rangle \langle 0| \otimes \biggl\{ \hat
\rho_s(t) + \gamma \int_t^{t + \Delta t} dt_1 \langle \hat n \rangle_{t_1}
\hat \rho_s(t) \nonumber \\
& & - \int_t^{t + \Delta t} dt_2 \int_t^{t_2} dt_1 \left[ e^{\frac{\gamma
(t_1 - t_2)}{2}} \left( \hat \sigma^\dagger \hat \sigma \hat \rho_s(t) +
\hat \rho_s(t) \hat \sigma^\dagger \hat \sigma \right) + \gamma^2 \langle
\hat n \rangle_{t_1} \langle \hat n \rangle_{t_2} \right] \biggr\}
\nonumber \\
& & + |1\rangle \langle 1| \otimes \int_t^{t + \Delta t} dt_2 \int_t^{t_2}
dt_1 2 e^{\frac{\gamma (t_1 + t_2)}{2}} \hat \sigma \hat \rho_s(t) \hat
\sigma^\dagger + \cdots,
\end{eqnarray}
where we have kept only relevant terms.
We take $\Delta t$ such that $\gamma \Delta t \gg 1$ and $\gamma \Delta t
A^2 / \gamma^2 \ll 1$ are satisfied.
From the last term of Eq.~(\ref{perturb}), we find that the mean photon
number is given by $\langle \hat n \rangle_{t + \Delta t} \simeq 4
\gamma^{-2} \langle \hat \sigma^\dagger \hat \sigma \rangle_t$, which is
$O(A^2 / \gamma^2)$.
The last term and the term including $\gamma^2 \langle \hat n
\rangle_{t_1} \langle \hat n \rangle_{t_2}$ in Eq.~(\ref{perturb}) are
therefore $O(A^2 / \gamma^2)$ and can be ignored.
Thus, Eq.~(\ref{perturb}) can be approximated as
\begin{equation} \label{spinmaster0}
\hat \rho(t + \Delta t) \simeq |0\rangle \langle 0| \otimes \left\{ \hat
\rho_s(t) - \frac{2 \Delta t}{\gamma} [\hat \sigma^\dagger \hat \sigma
\hat \rho_s(t) + \hat \rho_s(t) \hat \sigma^\dagger \hat \sigma] + \frac{4
\Delta t}{\gamma} \langle \hat \sigma^\dagger \hat \sigma \rangle_t \hat
\rho_s(t) \right\}.
\end{equation}
\end{widetext}
Since, the last two terms in the curly brackets in Eq.~(\ref{spinmaster0})
are $O(\gamma \Delta t A^2 / \gamma^2)$, which is assumed to be small,
we obtain an effective master equation of the spin state for the no-count
process:
\begin{equation} \label{SPINMASTER}
\frac{\partial \hat \rho_s(t)}{\partial t} = -\frac{2}{\gamma} [ \hat
\sigma^\dagger \hat \sigma \hat \rho_s(t) + \hat \rho_s(t) \hat
\sigma^\dagger \hat \sigma ] + \frac{4}{\gamma} \langle \hat
\sigma^\dagger \hat \sigma \rangle_t \hat \rho_s(t).
\end{equation}

It follows from Eq.~(\ref{SPINMASTER}) that the system evolves into the
lowest eigenstate of the operator
\begin{equation} \label{bb}
\hat \sigma^\dagger \hat \sigma = g^2 \hat S_+ \hat S_- + g A (\hat S_+ +
\hat S_-) + A^2,
\end{equation}
if the initial spin state is not orthogonal with the lowest eigenstate.
To see this, let us expand the density operator as
\begin{equation}
\hat \rho(t) = \sum_{n, m} c_{nm} |\phi_n \rangle \langle \phi_m |,
\end{equation}
where $|\phi_n \rangle$ is the eigenstate of the operator~(\ref{bb}) with
eigenvalue $\varepsilon_n$, i.e., $\hat \sigma^\dagger \hat \sigma |\phi_n
\rangle = \varepsilon_n |\phi_n \rangle$.
The master equation (\ref{SPINMASTER}) is rewritten as
\begin{equation}
\dot{c}_{nm}(t) = \frac{2}{\gamma} (2 \langle \hat \sigma^\dagger \hat
\sigma \rangle_t - \varepsilon_n - \varepsilon_m) c_{nm}(t),
\end{equation}
which can be solved to give
\begin{equation}
c_{nm}(t) = \exp\left[ \frac{4}{\gamma} \int^t dt' \langle \hat
\sigma^\dagger \hat \sigma \rangle_{t'} - \frac{2}{\gamma} (\varepsilon_n
+ \varepsilon_m) t \right] c_{nm}(0).
\end{equation}
Thus, the system develops into the state that minimizes $\varepsilon_n +
\varepsilon_m$, that is, the lowest eigenstate of the operator~(\ref{bb}),
if the corresponding initial coefficient $c_{nm}(0)$ is not zero.
We note that Eq.~(\ref{bb}) is independent of $\gamma$, and therefore
small experimental fluctuations in $\gamma$ do not affect the steady
state.
Since the relaxation coefficient in Eq.~(\ref{SPINMASTER}) is proportional
to $\gamma^{-1}$, the relaxation becomes slower for larger $\gamma$.
This is because the number of photons in the cavity is smaller for larger
$\gamma$, and therefore the state reduction caused by the no-count process
becomes less effective.

The steady state of the master equation (\ref{SPINMASTER}) is the SSS, as
will be numerically shown in Sec.~\ref{s:numerical}.
We evaluate perturbatively the degree of spin squeezing (\ref{SSS}) of the
steady state.
When $A \gg g$, the first-order perturbation with respect to $g / A$
yields (Appendix~\ref{a:perturb})
\begin{equation} \label{PERTURB1}
\zeta = 1 - \frac{g}{4A} (N - 1) < 1,
\end{equation}
showing that the state is an SSS.
When $A \ll g$, the second-order perturbation with respect to $A / g$
yields (Appendix~\ref{a:perturb})
\begin{equation} \label{PERTURB2}
\zeta = 1 - \frac{2 A^2}{N^2 g^2} < 1,
\end{equation}
showing that the state is again an SSS.

\subsection{Exact numerical results}
\label{s:numerical}

We numerically simulate the stochastic process of the continuous
measurement described by Eqs.~(\ref{onecount}) and (\ref{nocount1}) with
$\hat H_{\rm nc}$ given by Eq.~(\ref{atomnc}).
We divide the time into a small interval $\Delta t$ such that $\gamma
\langle \hat n \rangle_t \Delta t$ is always much smaller than unity.
We then generate a random number $r$ distributed uniformly between 0 and 1
for each time interval.
If $0 \leq r \leq \gamma \langle \hat n \rangle_t \Delta t$, we assume
that the one-count process occurs between $t$ and $t + \Delta t$ and
calculate the density operator at $t + \Delta t$ according to
Eq.~(\ref{onecount}).
If $\gamma \langle \hat n \rangle_t \Delta t < r \leq 1$, we assume that
the no-count process occurs during $t$ and $t + \Delta t$ and calculate
the density operator at $t + \Delta t$ according to Eq.~(\ref{nocount1}).

\begin{figure}[tb]
\includegraphics[width=8.4cm]{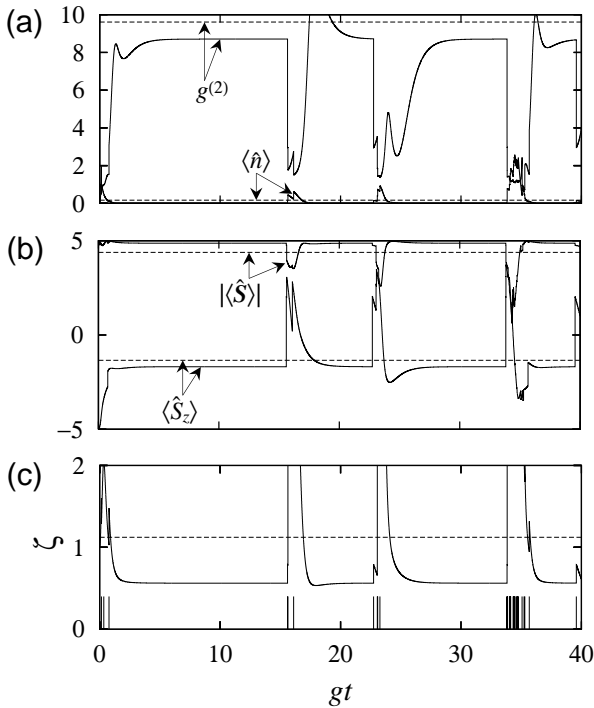}
\caption{
The solid curves show examples of time evolutions of $\langle \hat n
\rangle$, $g^{(2)} \equiv \langle \hat a^{\dagger 2} \hat a^2 \rangle /
\langle \hat n \rangle$, $\langle \hat S_z \rangle$, $|\langle \hat{\bf S}
\rangle|$, and $\zeta$ as defined in Eq.~(\protect\ref{SSS}) under
continuous photodetection with $N = 10$, $A = 5 g$, and $\gamma = 10 g$.
The initial state of the photon field is the vacuum state, and that of the
atoms is $|S, S_z = -S \rangle$.
The impulses in (c) indicate the times at which photons are detected.
The dashed lines show the values for the steady state in the non-referring
measurement.
}
\label{f:stochastic}
\end{figure}

Figure \ref{f:stochastic} shows an example of the stochastic time
evolutions of various quantities in the referring measurement process with
$A / g = 5$, $\gamma / g = 10$, and $N = 10$, where the photon field is
initially in the vacuum state and the atoms are in $|S, -S \rangle$.
The impulses that can be seen at the bottom of Fig.~\ref{f:stochastic} (c)
indicate the times at which photons are detected.
They are clearly bunched, reflecting the fact that $g^{(2)} \equiv \langle
\hat a^{\dagger 2} \hat a^2 \rangle / \langle \hat n \rangle^2$ in the
cavity is larger than unity most of the time as shown in
Fig.~\ref{f:stochastic} (a).
The bunching nature of photons makes the duration of each no-count process
longer, helping the system reach the no-count steady state (shown as
plateaus in Fig.~\ref{f:stochastic}).
We note that in the no-count steady state, the SSS is generated with
$\zeta \simeq 0.56$ [solid curve in Fig.~\ref{f:stochastic} (c)], where
$\zeta$ is defined in Eq.~(\ref{SSS}).
In the non-referring measurement process, in contrast, no SSS is obtained
[dashed line in Fig.~\ref{f:stochastic} (c), obtained by numerically
integrating Eq.~(\ref{master})], indicating that the information
concerning the detection times of photons is crucial for obtaining the
SSS.
It should be emphasized that in such a referring measurement process
$\zeta$ is smaller than unity most of the time, and therefore the SSS is
obtained as a quasi-steady state.

\begin{figure}[tb]
\includegraphics[width=8.4cm]{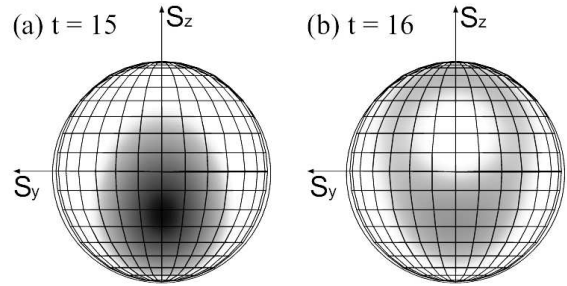}
\caption{
Gray-scale images of the quasi-probability distributions of the spin
state $\langle \theta, \phi | \hat \rho_s | \theta, \phi \rangle$, where
$\hat \rho_s$ is the reduced density operator of spin and $| \theta,
\phi \rangle$ is the coherent spin state defined in
Eq.~(\protect\ref{css}).
(a) At $t = 15$ in Fig.~\protect\ref{f:stochastic} (no-count steady state)
and (b) at $t = 16$ (just after a few photons were counted).
}
\label{f:sphere}
\end{figure}

Figure~\ref{f:sphere} (a) shows the quasi-probability distribution (QPD)
corresponding to the spin state at $t = 15$, when the no-count steady
state is reached.
We can see that the spin fluctuation is squeezed in the azimuthal
direction.
However, at $t = 16$ the spin state is disturbed by the photon counts and
the corresponding QPD shown in Fig.~\ref{f:sphere} (b) exhibits a pattern
similar to a rotated Dicke state.

\begin{figure}[tb]
\includegraphics[width=8.4cm]{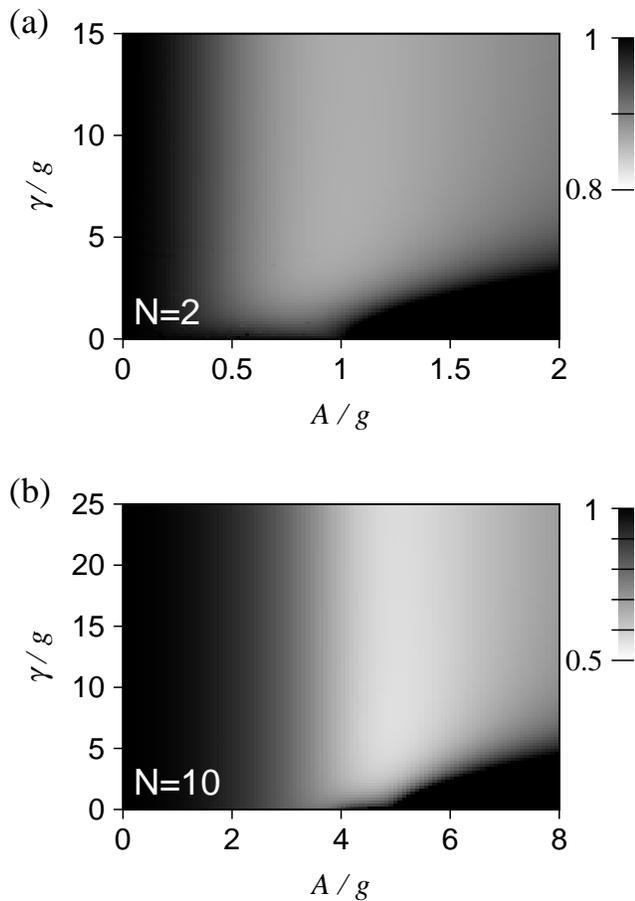}
\caption{
The dependence of the squeezing parameter $\zeta_{\rm st}$ on $A$ and
$\gamma$ for the no-count steady state with (a) $N = 2$ and (b) $N = 10$.
The legends on the right show the gray scale of the value of $\zeta_{\rm
st}$.
}
\label{f:contour}
\end{figure}

The degree of spin squeezing, $\zeta$, depends on the parameters $A$,
$\gamma$, and $N$.
Figure~\ref{f:contour} shows the dependence of $\zeta_{\rm st}$ on $A$ and
$\gamma$ for $N = 2$ and $10$, where $\zeta_{\rm st}$ is the steady-state
value of $\zeta$ in the no-count process.
We see that the SSS is obtained for a wide range of parameters.
We find that in both cases of $N = 2$ and $N = 10$ the maximum squeezing
is obtained for $A / g \simeq N / 2 = S$, and we have numerically
confirmed that this finding holds true for other values of $N$.
For $\gamma \gg A$, $\zeta_{\rm st}$ depends little on $\gamma$, in
agreement with the analytic results of Eqs.~(\ref{bb}), (\ref{PERTURB1}),
and (\ref{PERTURB2}).

\begin{figure}[tb]
\includegraphics[width=8.4cm]{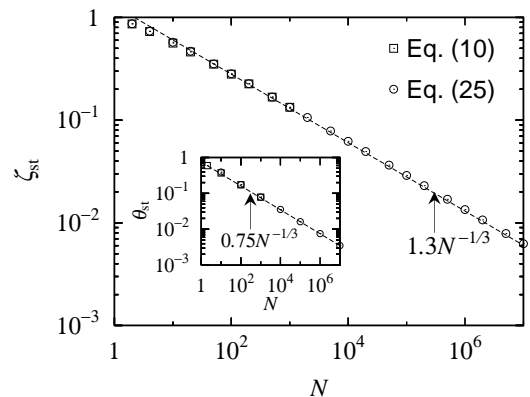}
\caption{
The $N$ dependence of the minimum squeezing parameter $\zeta_{\rm st}$ and
the corresponding spin angle $\theta_{\rm st} = \tan^{-1} (\langle \hat
S_z \rangle / \langle \hat S_x \rangle)$ (inset) of the no-count steady
state, where $\langle \hat S_x \rangle < 0$, $\langle \hat S_y \rangle =
0$, and $\langle \hat S_z \rangle < 0$.
Squares were obtained by solving the master equation (\ref{nocount1})
and circles by diagonalizing Eq.~(\ref{bb}).
}
\label{f:squeeze}
\end{figure}

Figure~\ref{f:squeeze} shows the dependence of $\zeta_{\rm st}$ and the
angle of the spin vector $\theta_{\rm st}$ on $N$, where the squares are
obtained by solving the master equation (\ref{nocount1}) and the circles
by diagonalizing Eq.~(\ref{bb}).
Both plots are in excellent agreement, confirming the validity of the
analytical treatment in Sec.~\ref{s:analytic}.
It is interesting to note that both $\zeta_{\rm st}$ and $\theta_{\rm st}$
are proportional to $N^{-1 / 3}$.
The other models of spin squeezing also exhibit such power laws: $N^{-2 /
3}$ for one-axis twisting and $N^{-1}$ for two-axis counter
twisting~\cite{Kitagawa}.
In a high-Q cavity, the Jaynes-Cummings interaction between photons and
atoms generates spin squeezing proportional to $N^{-1 / 4}$~\cite{SaitoA},
demonstrating that the present scheme is more effective for larger $N$.
Table I summarizes the dependence of $\zeta$ on $N$ in these schemes.
\begin{table}
\caption{
Power laws of the optimal squeezing parameter $\zeta$ with respect to the
number of atoms $N$ for various squeezing schemes.
}
\label{t:squeeze}
\begin{ruledtabular}
\begin{tabular}{cc}
 & $\zeta$ \\ \hline
two-axis twisting~\cite{Kitagawa} & $N^{-1}$ \\
one-axis twisting~\cite{Kitagawa} & $N^{-2/3}$ \\
Jaynes-Cummings (with measurement) & $N^{-1/3}$ \\
Jaynes-Cummings (without measurement)~\cite{SaitoA} & $N^{-1/4}$ \\
\end{tabular}
\end{ruledtabular}
\end{table}

\section{Discussions and Conclusions}
\label{s:conclusion}

When the number of excited atoms is much smaller than the total number of
atoms $N$, the Holstein-Primakoff transformation~\cite{Holstein} can be
approximated by $\hat S_- \simeq (2 S)^{1/2} \hat b$ and $\hat S_+ \simeq
(2 S)^{1/2} \hat b^\dagger$, where $\hat b$ and $\hat b^\dagger$ are the
bose operators satisfying $[\hat b, \hat b^\dagger] = 1$.
The non-Hermitian operator describing the no-count process (\ref{atomnc})
then becomes
\begin{equation} \label{bosonH}
\hat H_{\rm nc} / \hbar = A (\hat a + \hat a^\dagger) + G (\hat a \hat
b^\dagger + \hat a^\dagger \hat b) - \frac{i \gamma}{2} \hat n,
\end{equation}
where $G \equiv (2 S)^{1/2} g$.
In a similar manner to the derivation given in Appendix~\ref{a:pump}, one
can show that during the no-count process the system develops into $|0
\rangle_a | \beta \rangle_b$, where $|0 \rangle_a$ is the vacuum state of
the photon field and $| \beta \rangle_b = e^{\beta (\hat b^\dagger - \hat
b)} |0 \rangle_b$ is the coherent state of the boson with $\beta = -A /
G$.
Therefore, the bosonization of spin eliminates the squeezing properties of
atoms, and thus the spin algebra, which plays a role in nonlinear
saturation, is crucial for obtaining squeezing in the present scheme.

Experimentally, the model presented may be implemented in various systems,
such as a microwave cavity with Rydberg atoms~\cite{Haroche}, an optical
cavity with cold atoms~\cite{Kimble}, and trapped ions coupled with their
center-of-mass motion~\cite{Wineland}.
For example, let us consider a system in which Rb atoms in the circular
Rydberg state pass through a superconducting microwave
cavity~\cite{Haroche}.
The coupling constant between the atoms and the cavity photons can be
varied from 0 to 24 kHz~\cite{Haroche}.
If we take $\kappa / (2\pi) = 2$ kHz, $g / (2\pi) = 100$ Hz, $\tau_{\rm p}
= 10$ $\mu{\rm s}$, and $N = 10$, the conditions~(\ref{condition}) are
satisfied as $\kappa \tau_{\rm p} \simeq 0.1$ and $g S \tau_{\rm p} \simeq
A \tau_{\rm p} \simeq 0.03$, where $\kappa$ is the coupling constant
between probe atoms and photons, and $\tau_{\rm p}$ is the passage time
of a probe atom through the cavity (see Appendix~\ref{a:continuous}).
These parameters give $\gamma / g = \kappa^2 \tau_{\rm p} / g \simeq
2.5$, and the lifetime of the Rydberg state $\simeq 30$ ms~\cite{Haroche}
imposes the constraint $gt < 19$.
The cavity relaxation time now reaches 0.3 s~\cite{Walther}.
Thus the scheme is experimentally feasible.

In conclusion, we have studied the system in which $N$ two-level atoms
interact with photons in a pumped cavity, and have shown that the atomic
state develops into an SSS by continuous measurement of photon number in
the cavity.
In particular, the SSS is obtained as a quasi-steady state during the
no-count process.
The present method does not require fine tuning of the initial state and
the time and strength of interaction, and therefore significantly
facilitates experimental realization of the SSS.

\begin{acknowledgements}
This work was supported by the Special Coordination Funds for Promoting
Science and Technology and a Grant-in-Aid for Scientific Research (Grant
No. 11216204) by the Ministry of Education, Science, Sports, and Culture
of Japan, by the Yamada Science Foundation, and by the Toray Science
Foundation.
\end{acknowledgements}

\appendix

\section{A model of the continuous measurement of photon number}
\label{a:continuous}

For continuous measurement of photon number, the Srinivas-Davies
``ansatz''~\cite{Srinivas} is usually employed.
However, it is by no means obvious whether it is valid in our system, and
therefore, we explicitly derive nonunitary time evolutions
(\ref{onecount}) and (\ref{nocount2}) using the continuous photodetection
model of Ref.~\cite{Imoto}.

We consider a situation in which probe atoms prepared in the ground
state $|{\rm g} \rangle$ pass through the cavity successively, and the
internal states of the atoms coming out of the cavity are detected by,
e.g., an ionization detector (Fig.~\ref{f:system}).
When a probe atom is detected to be in the excited state $|{\rm e}
\rangle$, we regard this event as a one-count process.
When the atom coming out of the cavity remains in the ground state, we
regard this event as a no-count process.

The time evolution as the probe atom passes through the cavity is
described by the Hamiltonian
\begin{equation}
\hat H_{\rm p} = \hat H_0^{\rm rot} + \hbar \kappa (\hat a \hat \sigma_+
+ \hat a^\dagger \hat \sigma_-),
\end{equation}
where $\hat H_0^{\rm rot}$ is given in Eq.~(\ref{Hrot}), $\hat \sigma_+$
and $\hat \sigma_-$ are the raising and lowering operators of the probe
atom, and $\kappa$ is a coupling constant which we assume to be real.
We assume that the transition frequency of the probe atom is resonant with
the cavity mode.
When the probe atom enters the cavity at time $t$ and comes out of it at
$t + \tau_{\rm p}$, the density operator $\hat \rho_{\rm sp}$ for the
system and the probe atom evolves as
\begin{eqnarray} \label{perturba}
\hat \rho_{\rm sp}(t + \tau_{\rm p}) & = & \hat \rho(t) \otimes |{\rm g}
\rangle \langle {\rm g}| - \frac{i \tau_{\rm p}}{\hbar} [\hat H_{\rm p},
\hat \rho(t) \otimes |{\rm g} \rangle \langle {\rm g}|] \nonumber \\
& & - \frac{\tau_{\rm p}^2}{2 \hbar^2} [\hat H_{\rm
p} [\hat H_{\rm p}, \hat \rho(t) \otimes |{\rm g} \rangle \langle {\rm
g}|]] + \cdots,
\end{eqnarray}
where $\hat \rho(t)$ is the density operator for the system (excluding the
probe atom).

Here we assume that $\tau_{\rm p}$ is sufficiently small so that third
and higher-order terms in Eq.~(\ref{perturba}) can be ignored.
This approximation is valid if the change in the state of the system
is small during $\tau_{\rm p}$.
This condition is satisfied if the following conditions are met:
\begin{equation} \label{condition}
A \tau_{\rm p}^{-1} \ll 1, \;\;\; g S \tau_{\rm p}^{-1} \ll 1, \;\;\;
\delta_{\rm cp} \tau_{\rm p}^{-1} \ll 1, \;\;\; \delta_{\rm ap} \tau_{\rm
p}^{-1} \ll 1.
\end{equation}

When the probe atom is detected to be in the excited state $| {\rm e}
\rangle$, the density operator of the system is projected into
\begin{equation} \label{projecte}
{\rm Tr}_{\rm probe}[| {\rm e} \rangle \langle {\rm e} | \hat \rho_{\rm
sp}(t + \tau_{\rm p})] = \kappa^2 \tau_{\rm p}^2 \hat a \hat \rho(t) \hat
a^\dagger.
\end{equation}
The normalized form of Eq.~(\ref{projecte}) gives the density operator
immediately after the no-count process:
\begin{equation}
\hat\rho(t_+) = \frac{\hat a \hat\rho(t) \hat a^\dagger}{{\rm Tr}[\hat a
\hat\rho(t) \hat a^\dagger]},
\end{equation}
where $t_+$ denotes a time infinitesimally later than $t$.
The probability that the one-count process (\ref{projecte}) occurs is
given by the trace of Eq.~(\ref{projecte}), i.e., $\gamma \tau_{\rm p}
\langle \hat n \rangle_t$, where
\begin{equation} \label{gamma}
\gamma \equiv \kappa^2 \tau_{\rm p}
\end{equation}
can be interpreted as the rate of photodetection per one photon.

When the probe atom is detected in its ground state $| g \rangle$, the
density operator of the system is projected into
\begin{eqnarray}
& & {\rm Tr}_{\rm probe}[| {\rm g} \rangle \langle {\rm g} | \hat
\rho_{\rm sp}(t + \tau_{\rm p})] \nonumber \\
& \simeq & \hat \rho(t) - \frac{i
\tau_{\rm p}}{\hbar} [\hat H_0^{\rm rot}, \hat \rho(t)] - \frac{\tau_{\rm
p}^2}{2 \hbar^2} [\hat H_0^{\rm rot}, [\hat H_0^{\rm rot}, \hat \rho(t)]
\nonumber \\
& & - \frac{\tau_{\rm p}^2 \kappa^2}{2} \left[ \hat n \hat \rho(t) + \hat
\rho(t) \hat n \right]
\nonumber \\
& \simeq & e^{-\tau_{\rm p} \left(\frac{i}{\hbar} \hat H_0^{\rm rot} -
\frac{\gamma}{2}\right)} \hat \rho(t) e^{\tau_{\rm p} \left(
\frac{i}{\hbar} \hat H_0^{\rm rot} - \frac{\gamma}{2} \right)}.
\end{eqnarray}
When successive probe atoms are found to be in the ground state
over a time interval $T$, the density operator of the system after such
a no-count process is given by Eq.~(\ref{nocount2}).

Probe atoms were introduced to explicitly derive Eqs.~(\ref{onecount}) and
(\ref{nocount2}).
However, these expressions may also describe a more general system under
continuous photodetection, such as the situation in which photons leaking
out of the cavity are detected by a photodetector.

\section{Steady state in a pumped cavity under continuous photodetection}
\label{a:pump}

We show that the state of the system described by the Hamiltonian
(\ref{Hrot}) eventually evolves into a coherent state in both referring
and non-referring measurements.

The non-Hermitian operator that describes the no-count process
(\ref{nocount2}) is rewritten in the rotating frame by
\begin{eqnarray}
\hat H_{\rm nc} / \hbar & = & A (\hat a + \hat a^\dagger) + (\delta_{\rm
cp} - i \gamma / 2) \hat n \nonumber \\
& = & \Omega \left( \hat a^\dagger + \frac{A}{\Omega} \right) \left( \hat
a + \frac{A}{\Omega} \right) - \frac{A^2}{\Omega} \nonumber \\
& = & \Omega  e^{\frac{A}{\Omega}(\hat a + \hat a^\dagger)} \hat n
e^{-\frac{A}{\Omega}(\hat a + \hat a^\dagger)} - \frac{A^2}{\Omega},
\end{eqnarray}
where we define $\Omega \equiv \delta_{\rm cp} - i \gamma / 2$.
We then obtain
\begin{eqnarray} \label{app1eq}
e^{-i \hat H_{\rm nc} t / \hbar} & \propto &
e^{\frac{A}{\Omega}(\hat a + \hat a^\dagger)} e^{-i \Omega t \hat n}
e^{-\frac{A}{\Omega}(\hat a + \hat a^\dagger)} \nonumber \\
& = & \sum_{n = 0}^\infty e^{-i n \delta_{\rm cp} t - n \gamma t / 2}
e^{\frac{A}{\Omega}(\hat a + \hat a^\dagger)} |n \rangle \langle n|
e^{-\frac{A}{\Omega}(\hat a + \hat a^\dagger)}. \nonumber \\
\end{eqnarray}
For $\gamma t \gg 1$, the term corresponding to $n = 0$, i.e.,
$e^{\frac{A}{\Omega}(\hat a + \hat a^\dagger)} |0 \rangle \langle 0|
e^{-\frac{A}{\Omega}(\hat a + \hat a^\dagger)} \propto |\alpha \rangle
\langle \alpha^*|$ becomes dominant in the summation, where $|\alpha
\rangle$ is the coherent state with $\alpha = -A / \Omega$.
We thus obtain
\begin{equation} \label{coherent}
\lim_{t \rightarrow \infty} \frac{e^{-i \hat H_{\rm nc} t / \hbar}
|\psi \rangle}{||e^{-i \hat H_{\rm nc} t / \hbar} |\psi \rangle||} = 
|\alpha \rangle
\end{equation}
for $\langle \alpha^* | \psi \rangle \neq 0$.
Therefore, during the no-count process (\ref{nocount2}) the state develops
into $|\alpha \rangle$ unless the initial state is orthogonal to
$|\alpha^* \rangle$.
If $\langle \alpha^* | \psi \rangle = 0$, the term of $n = 1$ in the
summation of Eq.~(\ref{app1eq}) becomes dominant, and the state develops
into a displaced number state.

If the photons are detected at times $t_1, t_2, \cdots, t_n$, the time
development operator (\ref{refP}) for the referring measurement takes the
form
\begin{eqnarray} \label{P}
& & \hat P(T; t_1, \cdots, t_n) \nonumber \\
& = & e^{-i \hat H_{\rm nc} T / \hbar} \Pi_{k = 1}^n \left[ e^{-i \Omega
t_k} \hat a + \frac{A}{\Omega} (e^{-i \Omega t_k} - 1) \right].
\end{eqnarray}
From Eqs.~(\ref{coherent}) and (\ref{P}) we obtain
\begin{equation} \label{app1eq2}
\lim_{T \rightarrow \infty} \hat P(T; t_1, \cdots, t_n) |\psi \rangle
\propto |\alpha \rangle.
\end{equation}
Thus, the state develops to $|\alpha \rangle$ in the referring
measurement.
This result is also true for non-referring measurement, since
Eq.~(\ref{app1eq2}) depends neither on the number of detected photons nor
on the times at which photons are detected.

\section{Proof of Eqs.~(\protect\ref{PERTURB1}) and
(\protect\ref{PERTURB2})}
\label{a:perturb}

We calculate perturbatively~\cite{Sakurai} the squeezing parameter for the
lowest eigenstate of the operator~(\ref{bb}).

When $g / A \ll 1$, the first term on the rhs of Eq.~(\ref{bb}) can be
treated as a perturbation.
For convenience, we rotate in the spin space by $\pi / 2$ with respect to
the $S_y$ axis.
The operator~(\ref{bb}) then becomes
\begin{equation} \label{bbrot}
\hat \sigma^\dagger \hat \sigma = A^2 \left[1 + \frac{2g}{A} \hat S_z +
\frac{g^2}{A^2} (\hat {\bf S}^2 - \hat S_x^2 - \hat S_x) \right].
\end{equation}
The unperturbed ground state is then taken to be the eigenstate of $\hat
S_z$ with eigenvalue $-S$, which we write as $|S, -S \rangle$.
Performing first-order perturbation with respect to the
last term of Eq.~(\ref{bbrot}), we obtain the ground state as
\begin{equation}
|S, -S \rangle + \frac{\sqrt{2S} g}{4A} \left( |S, 1 - S \rangle +
\frac{1}{4} \sqrt{2 (2S - 1)} |S, 2 - S \rangle \right),
\end{equation}
with the squeezing parameter calculated to be
\begin{equation}
\zeta = 1 - \frac{g}{4A} (2S - 1).
\end{equation}

When $A \ll g$, the second term on the rhs of Eq.~(\ref{bb}) may be
treated as a perturbation.
In this case, we must perform second-order perturbation to obtain spin
squeezing.
The ground state is then given by
\begin{eqnarray}
& & \left( 1 - \frac{A^2}{4S g^2} \right) |S, -S \rangle - \frac{A}{\sqrt{2S}
g}  |S, 1 - S \rangle \nonumber \\
& & + \frac{A^2}{g^2} \frac{1}{\sqrt{4S (2S - 1)}} |S, 2
- S \rangle,
\end{eqnarray}
with the squeezing parameter calculated to be
\begin{equation}
\zeta = 1 - \frac{A^2}{2S^2 g^2}.
\end{equation}

\end{document}